\documentclass[prc,showpacs,preprintnumbers,amsmath,amssymb]{revtex4}
\usepackage{graphicx}
\newcommand{\beq}{\begin{equation}}
\newcommand{\eeq}{\end{equation}}
\newcommand{\beqa}{\begin{eqnarray}}
\newcommand{\eeqa}{\end{eqnarray}}
\newcommand{\nn}{\nonumber}
\newcommand{\Sigs}{\Sigma_{\mathrm s} }
\newcommand{\Sigv}{\Sigma_{\mathrm v} }
\newcommand{\Sigo}{\Sigma_{\mathrm o} }
\newcommand{\kf}{k_{\mathrm F} }

\newcommand{\kfj}{k_{\mathrm Fj} }
\newcommand{\bfgamma}{\mbox{\boldmath$\gamma$\unboldmath}}
\newcommand{\veck}{\textbf{k}}

\newcommand{\vecq}{\textbf{q}}

\begin{document}
\preprint{}
\title{Off-Shell Behavior of Nucleon Self-Energy in Asymmetric Nuclear
Matter}
% in the Framework of a Relativistic DBHF Approach.}
\author{E. N. E. van Dalen}
\email{eric.van-dalen@uni-tuebingen.de}
\author{H. M\"uther}
\affiliation{Institut f$\ddot{\textrm{u}}$r Theoretische Physik,
Universit$\ddot{\textrm{a}}$t T$\ddot{\textrm{u}}$bingen, Auf der
Morgenstelle 14, D-72076 T$\ddot{\textrm{u}}$bingen, Germany}
\begin{abstract}
The off-shell behavior of the nucleon self-energy in isospin asymmetric nuclear 
matter is investigated within the framework of relativistic
Dirac-Brueckner-Hartree-Fock approach based on projection techniques. The
dependence of the Dirac components of the self-energy on momentum as well as
energy is evaluated for symmetric as well as asymmetric nuclear matter. Special
attention is paid to the various contributions to the momentum dependence of the
real and imaginary part of the optical potential. The consequences to the
different definitions of the effective nucleon mass and particle spectral
functions are discussed.
\end{abstract}
\maketitle
%%%%%%%%%%%%%%%%%%%%%%%%%%%%%%%%%%%%%%%%%%%%%%%%%%%%%%%%%%%%%%%%%%%%%%%%
\section{Introduction}

The investigation of isospin asymmetric nuclear matter is receiving a lot of
attention as the exploration of nuclear systems outside the valley of stable
nuclei are of high interest for
astrophysical and nuclear structure studies. In the field of astrophysics
these investigations are important for the physics of supernova
explosions~\cite{bethe:1990} and of neutron
stars~\cite{pethick:1995,vandalen:2003,gogelein:2008}, whereas in the field
of nuclear structure it is of interest in the study of  neutron-rich
nuclei~\cite{tanihata:1995,hansen:1995}. The new generation of radioactive
beam facilities, e.g. the future GSI facility FAIR in Germany or SPIRAL2 at
GANIL/France, facilitate this kind of nuclear structure studies. Off-shell
effects are crucial to describe the data obtained from the 
collisions occurring in these radioactive beam experiments.

The theoretical models which make predictions on the equation of state (EoS)
of nuclear matter can roughly be divided in the following three classes:
Phenomenological density functionals, effective field theory (EFT)
approaches, and ab initio approaches. Phenomenological density functionals
are based on effective density dependent interactions with usually between
six and 15 parameters.  The effective field theory approaches lead to a more
systematic expansion of the EoS in powers of the Fermi momentum $k_F$,
respectively the density, with a small number of free parameters. The
parameters of these models are typically adjusted to reproduce the
properties of normal nuclei. Therefore extrapolations outside the valley of
stable nuclei must be considered with some scepticism. 

Ab initio
approaches, such as the Brueckner-Hartree-Fock (BHF) and the
Dirac-Brueckner-Hartree-Fock (DBHF) approach, are based on high precision
free space nucleon-nucleon interactions and the nuclear many-body problem is
treated microscopically. These approaches are more ambitious, since the
predictions for the nuclear EoS are essentially parameter free. Therefore,
they have a higher predictive power also for exotic nuclear systems. 

Although non-relativistic ab-initio calculations were able to describe the
nuclear saturation mechanism qualitatively, they failed quantitatively.
Three-body forces were included in these non-relativistic microscopic
calculations to fit the empirical saturation point of symmetric nuclear
matter as well as the properties of light nuclei. A major breakthrough was
achieved when the first relativistic DBHF calculations were
performed~\cite{anastasio:1983,horowitz:1987,brockmann:1990}. It could
describe the saturation properties of nuclear matter without any need to
introduce a three-nucleon force. In fact, it has been argued that the
three-nucleon forces required in non-relativistic calculations have to be
introduced to simulate the change of the Dirac spinors in the nuclear
medium, which is contained in relativistic calculations~\cite{brown:1987}.

Beside this success of predicting the empirical saturation point also the
strength of the spin-orbit term in the single-particle spectrum of finite
nuclei and the momentum dependence of the optical potential for
nucleon-nucleus scattering\cite{mahaux:1980,kleinm:1994} were considered as
fingerprints of relativistic effects in nuclear structure physics at low
energies.  

However, relativistic microscopic DBHF investigations of isospin asymmetric
nuclear matter are rather
rare~\cite{ulrych:1997,dejong:1998,alonso:2003,vandalen:2004b,vandalen:2007}.
Furthermore, all these studies are restricted to the on-shell behavior of
nucleon properties in contrast to some  microscopic non-relativistic
investigations, which do include the study of off-shell behavior of these
properties in isospin asymmetric nuclear
matter~\cite{zuo:1999,hassaneen:2004}. Only in isospin symmetric nuclear
matter some attention is also paid to off-shell behavior in the framework of
relativistic microscopic studies~\cite{dejong:1991,dejong:1996}. This means
that in microscopic relativistic frameworks the off-shell behavior of
nucleon properties in isospin asymmetric nuclear matter has not been
investigated so far.

In this work we describe the off-shell behavior of nucleon properties in
isospin asymmetric nuclear  matter in the relativistic DBHF approach using
the Bonn A potential and its bare $NN$ matrix elements
$V$~\cite{machleidt:1989}. Furthermore, the optimal representation scheme
for the $T$-matrix, the subtracted $T$-matrix representation, is applied. In
this framework, the dependence of the off-shell behavior of nucleon
properties on the nuclear asymmetry is explored. Properties considered are
the optical potential, spectral functions, single-particle energies, and
masses. Our predictions will be compared to those of non-relativistic
calculations. Quantities of special interest are  the $k$-mass and the
$E$-mass, since   a rigorous  distinction between these two masses can only
be obtained from the knowledge of the off-shell behavior of the optical
potential. 

The plan of this paper is as follows. The relativistic DBHF approach is
discussed in Sec.~\ref{sec:DBHFA}. Furthermore, Sec.~\ref{sec:CR&SE} is
devoted to the covariant representation of the in-medium $T$-matrix in
connection with the nucleon self-energy components depending on energy and
momentum. The results are presented and discussed in Sec.~\ref{sec:R}.
Finally, we end with a summary and a conclusion in Sec.~\ref{sec:S&C}.

\section{DBHF approach}
\label{sec:DBHFA}
In this section the relativistic Brueckner approach is discussed. The
approach is roughly based on the ones in
Ref.~\cite{vandalen:2004b,vandalen:2007},   with the exception of some
modifications to separate the momentum and energy dependence. First a
general overview is given, followed by a discussion of the modifications.

In the relativistic Brueckner approach, the in-medium interaction of the
nucleons is treated in the ladder approximation of the relativistic
Bethe-Salpeter (BS) equation:
\beqa
T = V + i \int  V Q G G T,
\label{subsec:SM;eq:BS}
\eeqa
where $T$ denotes the $T$-matrix, $V$ is the bare nucleon-nucleon
interaction, and $Q$ is the Pauli operator. The Green's function $G$
describes the propagation of dressed nucleons in nuclear matter.
Furthermore, it fulfills the Dyson equation:
\beqa
G=G_0+G_0\Sigma G,
\label{subsec:SM;eq:Dysoneq}
\eeqa
where $G_{0}$ is the free nucleon propagator.
The self-energy $\Sigma$ in the
Hartree-Fock approximation is given by
\beqa
\Sigma = -i \int\limits_{F} (Tr[G T] - GT ).
\label{subsec:SM;eq:HFselfeq1}
\eeqa
The coupled set of
Eqs.~(\ref{subsec:SM;eq:BS})-(\ref{subsec:SM;eq:HFselfeq1})
represents a self-consistency problem and has to be iterated until
convergence is reached. 

To solve the self-consistency problem some approximations have to be made in
the iteration procedure. The first one is the quasi-particle approximation, 
i.e. the imaginary part of the self-energy $\Im m \Sigma$ will be neglected.
In addition, the ``reference spectrum approximation''~\cite{bethe:1963},
i.e. the effective mass of the nucleon is assumed to be entirely density
dependent ($|\veck|=k_F$), is applied. Furthermore, the two-particle
propagator $iGG$ in the BS equation is replaced by the  Thompson propagator
to reduce the four-dimensional BS integral equation,
Eq.~(\ref{subsec:SM;eq:BS}), to the three-dimensional Thompson equation.
After a partial wave projection onto the $|JMLS>$-states this Thomas
equation reduces to a set of one-dimensional integral equations over the
relative momentum. To achieve this reduction to the one-dimensional integral
equations the Pauli operator $Q$ is replaced by an angle-averaged Pauli
operator $\overline{Q}$ ~\cite{horowitz:1987}. For more details we refer to
\cite{vandalen:2004b,vandalen:2007}.

At the end of the iteration procedure, we keep the explicit momentum and
energy dependence in contrast to Ref~\cite{vandalen:2004b,vandalen:2007},
in which the starting energy is replaced by its on-shell value. In this
way, one obtains the nucleon self-energy   
\beqa
\Sigma(|\veck|,\omega)= \Sigs (|\veck|,\omega) -\gamma_0 \, \Sigo (|\veck|,\omega) +
\bfgamma  \cdot \textbf{k} \,\Sigv (|\veck|,\omega),
\label{subsec:SM;eq:self1}
\eeqa
as a function of the absolute momentum $|\veck|$ and energy $\omega$. These
self-energy components contain apart from a real part also an imaginary
part, which also can be calculated at the end of the iteration procedure.
These components of the self-energy are easily determined by taking the
respective traces \cite{horowitz:1987,sehn:1997}
\beqa
\Sigs = \frac{1}{4} tr \left[ \Sigma \right],\quad
\Sigo = \frac{-1}{4} tr \left[ \gamma_0 \, \Sigma \right], \quad
\Sigv =  \frac{-1}{4|\veck|^2 }
tr \left[{\bfgamma}\cdot \veck \, \Sigma \right].
\label{subsec:SM;eq:trace}
\eeqa
The other quantities such as the effective Dirac mass, single-particle
energy, and the optical potential can be obtained from these self-energy
components. 

\section{Covariant representation and the self-energy components}
\label{sec:CR&SE}

Since the $T$-matrix elements are determined in the two-particle c.m.
frame, a representation with covariant operators and Lorentz invariant
amplitudes in Dirac space is the most convenient way to Lorentz-transform
the $T$-matrix from the two-particle c.m. frame into  the nuclear matter
rest frame \cite{horowitz:1987}. However, some freedom in the choice of
this representation exists, because pseudoscalar ($ps$) and pseudovector
($pv$) components can not uniquely be disentangled. This ambiguity is
minimized by using the subtracted $T$-matrix representation scheme.
Therefore, the single-$\pi$ and-$\eta$ exchange are separated from the
full $T$-matrix. The contributions stemming from the single-$\pi$
and-$\eta$ exchange are given in the complete $pv$ representation, whereas
for the remaining part of the $T$-matrix,
\beq
T_{Sub}=T - V_{\pi,\eta},
\label{mixed1}
\eeq
the $ps$ representation is chosen. 

For the $ps$ representation the following set of five linearly independent
covariants 
\beqa
{\rm S}  =  1_1 \otimes 1_2, \\
{\rm V}  =  (\gamma^{\mu})_1 \otimes (\gamma_{\mu})_2, \\
{\rm T}  =  (\sigma^{\mu\nu})_1 \otimes (\sigma_{\mu\nu})_2, \\
{\rm A}  =  (\gamma_5)_1 (\gamma^{\mu})_1 \otimes (\gamma_5)_2 (\gamma_{\mu})_2,  \\
{\rm P}  =  (\gamma_5)_1 \otimes (\gamma_5)_2, 
\eeqa
are used in isospin symmetric nuclear matter. The interchanged invariants
are defined as~\cite{tjon:1985a}  $\tilde{{\rm
    S}}=\tilde{{\rm S}} {\rm S}$, $\tilde{{\rm
    V}}=\tilde{{\rm S}} {\rm V}$, $\tilde{{\rm
    T}}=\tilde{{\rm S}} {\rm T}$, $\tilde{{\rm
    A}}=\tilde{{\rm S}} {\rm A}$, and $\tilde{{\rm
    P}}=\tilde{{\rm S}} {\rm P}$ with operator $\tilde{{\rm
    S}}$ exchanging particles 1 and 2, i.e. $\tilde{{\rm
    S}} u(1)_{\sigma} u(2)_{\tau} = u(1)_{\tau} u(2)_{\sigma}$. 
In isospin asymmetric nuclear, one needs an additional covariant for the
np channel. It is defined as 
\beqa
{\rm I} = 1_1 \otimes (\gamma \cdot k)_2 + (\gamma \cdot k)_1 \otimes 1_2.
\eeqa
Taking the single nucleon momentum $\veck=(0,0,|\veck|)$ along the
$z$-axis, then we have for the self-energy components in the $ps$
representation scheme:
\beqa
\Sigma_s^{ij} (|\veck|,\omega) & = & 
\frac{1}{4}
\int_0 ^{\kfj} \frac{d^3\vecq}{(2 \pi)^3}  \frac{m^*_j}{E^*_{q,j}} 
[4 F^{ij}_{\rm S} - F^{ij}_{\rm \tilde{S}} - 4 F^{ij}_{\rm \tilde{V}} - 12
F^{ij}_{\rm \tilde{T}} + 4 F^{ij}_{\rm \tilde{A}} - F^{ij}_{\rm \tilde{P}} \nn \\ 
& &+  4 (1-\delta_{ij}) \frac{k^{* \mu} q^*_{\mu} - m_j^{*2}  }{m_j^*}   F^{ij}_{\rm I} ],
\label{subsec:PS;eq:s}
\eeqa
\beqa
\Sigma_o^{ij} (|\veck|,\omega) & = & 
 \frac{1}{4} 
\int_0^{\kfj} \frac{d^3\vecq}{(2 \pi)^3}  
[- 4 F^{ij}_{\rm V} + F^{ij}_{\rm \tilde{S}} - 2 F^{ij}_{\rm
    \tilde{V}}  - 2 F^{ij}_{\rm \tilde{A}} - F^{ij}_{\rm \tilde{P}} \nn \\ + & & 
4 (1-\delta_{ij}) m^*_j \frac{E^*_{k,i} - E^*_{q,j}}{E^*_{q,j}}  F^{ij}_{\rm I} ] ,
\label{subsec:PS;eq:o}
\eeqa
and
\beqa
\Sigma_v^{ij} (|\veck|,\omega) & = & 
\frac{1}{4} 
\int_0^{\kfj} \frac{d^3\vecq}{(2 \pi)^3} \frac{\vecq \cdot \veck}{|\veck|^2 E^*_{q,j}} 
[- 4 F^{ij}_{\rm V} + F^{ij}_{\rm \tilde{S}} - 2 F^{ij}_{\rm
    \tilde{V}}  - 2 F^{ij}_{\rm \tilde{A}} - F^{ij}_{\rm \tilde{P}} \nn \\ & & -4 (1-\delta_{ij}) m^*_j \frac{|\veck|-q_z}{q_z} F^{ij}_{\rm I} ],
\label{subsec:PS;eq:v}
\eeqa
where $k^{* \mu}_i=(E^*_{k,i},0,0,|\veck|)$. A relation exists between our
definition of the energy $\omega$ and $E^*_{k,i}=\omega+\Sigma_o^{i}
(|\veck|,\omega)+M$. Furthermore, the Lorentz invariant amplitudes $F$
have a dependence on as well the absolute momentum $|\veck|$ as the energy
$\omega$. 
In the complete $pv$ representation, one first applies the identities
\beqa
\frac{1}{2} ({\rm T} + {\rm \tilde{T}})= {\rm S} + {\rm
  \tilde{S}} + {\rm P}  + {\rm \tilde{P}}, 
\label{identi1}\\
{\rm V}  + {\rm \tilde{V}} = {\rm S} + {\rm \tilde{S}}  - {\rm
  P}  - {\rm \tilde{P}}
\label{identi2}
\eeqa
to replace tensor and vector covariants. Next the pseudoscalar covariant
${\rm P}=(\gamma_5)_1\otimes (\gamma_5)_2$ in the $T$-matrix
representation is replaced by the pseudovector covariant, 
\beqa
{\rm PV} = \frac{(\gamma_5 \gamma_{\mu})_1 p^{\mu}}{m^*_i+m^*_j} \otimes \frac{(\gamma_5 \gamma_{\mu})_2 p^{\mu}}{m^*_i+m^*_j},
\eeqa
with $p^{\mu}=k^{\mu}-q^{\mu}$. The contributions to the self-energy
components are then given by 
\beqa
\Sigma_s^{ij} (|\veck|,\omega) & = & 
\frac{1}{4}
\int_0^{\kfj} \frac{d^3\vecq}{(2 \pi)^3}  \frac{m^*_j}{E^*_{q,j}} 
[4 g^{ij}_{\rm S} - g^{ij}_{\rm \tilde{S}} + 4 g^{ij}_{\rm A} 
+ \frac{m_j^{*2} + m_i^{*2} - 2 k^{* \mu} q^*_{\mu}}{(m_i^*+m_j^*)^2} g^{ij}_{\rm \widetilde{PV}}
\nn \\  & & +  4 (1-\delta_{ij}) \frac{k^{* \mu} q^*_{\mu} - m_j^{*2}  }{m_j^*}   g^{ij}_{\rm I}],
\eeqa
\beqa
\Sigma_o^{ij} (|\veck|,\omega) & = & 
+ \frac{1}{4} \int_0^{\kfj} \frac{d^3\vecq}{(2 \pi)^3}  
[ g^{ij}_{\rm \tilde{S}} - 2 g^{ij}_{\rm A} -  \frac{2 E^*_{k,i} (m_j^{*2}-k^{* \mu}
q^*_{\mu}) - E^*_{q,j} (m^{*2}_j - m^{*2}_i)}{E^*_{q,j} (m^*_i +
    m^*_j)^2}   g^{ij}_{\rm \widetilde{PV}} \nn \\ & & +  4 (1-\delta_{ij}) m^*_j \frac{E^*_{k,i} - E^*_{q,j}}{E^*_{q,j}}  g^{ij}_{\rm I}],
\eeqa
and
\beqa
\Sigma_v^{ij} (|\veck|,\omega) & = & 
\frac{1}{4} 
\int_0^{\kfj} \frac{d^3\vecq}{(2 \pi)^3} \frac{\vecq \cdot \veck}{|\veck|^2 E^*_{q,j}} 
[g^{ij}_{\rm \tilde{S}} - 2 g^{ij}_{\rm A} - \frac{2 k^*_z (m^{*2}_j - k^{* \mu}
  q^*_{\mu} ) - q_z (m^{*2}_j - m^{*2}_i)}{q_z (m^*_i+m^*_j)^2} g^{ij}_{\rm \widetilde{PV}} \nn
\\ & & - 4 (1-\delta_{ij}) m^*_j \frac{|\veck|-q_z}{q_z} g^{ij}_{\rm I} ],
\eeqa
where the new amplitudes $g$ are defined as 
\beqa
\hspace{2cm}
\left( 
\begin{array}{c} 
g_{\rm S}^{{\rm ij}} \\ 
g_{\rm \tilde{S}}^{{\rm ij}} \\ 
g_{\rm A}^{{\rm ij}} \\ 
g_{\rm P}^{{\rm ij}} \\ 
g_{\rm \tilde{P}}^{{\rm ij}} \\
\end{array} 
\right)
= \frac{1}{4}
\left( 
{ 
\begin{array}{ccccccc} 
 4 & -2 & -8  & 0  & -2 \\
 0 & -6 & -16 & 0  &  2 \\
 0 & -2 &  0  & 0  & -2 \\
 0 &  2 & -8  & 4  &  2 \\
 0 &  6 & -16 & 0  & -2 \\
\end{array}} 
\right)
\left( 
\begin{array}{c} 
F_{\rm S}^{\rm ij} \\ 
F_{\rm V}^{\rm ij} \\ 
F_{\rm T}^{\rm ij} \\ 
F_{\rm P}^{\rm ij} \\ 
F_{\rm A}^{\rm ij} \\
\end{array} 
\right) 
\label{transform3}
\eeqa
and $g_{\rm I}^{{\rm ij}}=F^{ij}_{\rm I}$.  Finally, the total neutron and
proton self-energies including all channels can be written as
\beqa
\Sigma^n(|\veck|,\omega) = \Sigma^{nn}(|\veck|,\omega) + \Sigma^{np}(|\veck|,\omega) \quad; \Sigma^p(|\veck|,\omega) =\Sigma^{pp}(|\veck|,\omega) + \Sigma^{pn}(\veck,\omega),
\eeqa
respectively.

\section{Results}
\label{sec:R}
In the following we present the results for the off-shell properties of
isospin symmetric and asymmetric nuclear matter obtained from the  DBHF
approach based on projection techniques. The applied projection is the
subtracted $T$-matrix representation scheme.  Furthermore, the
nucleon-nucleon potential used is Bonn A. The presented results are
obtained from calculations performed at a density of n$_B$=0.181 fm$^{-3}$
in isospin symmetric nuclear matter and in isospin asymmetric nuclear
matter with the asymmetry parameter of $\beta = (n_n - n_p)/n_B=0.5$. 

\subsection{Self-energy}
\begin{figure}[!ht]
\begin{center}
\includegraphics[width=0.6\textwidth] {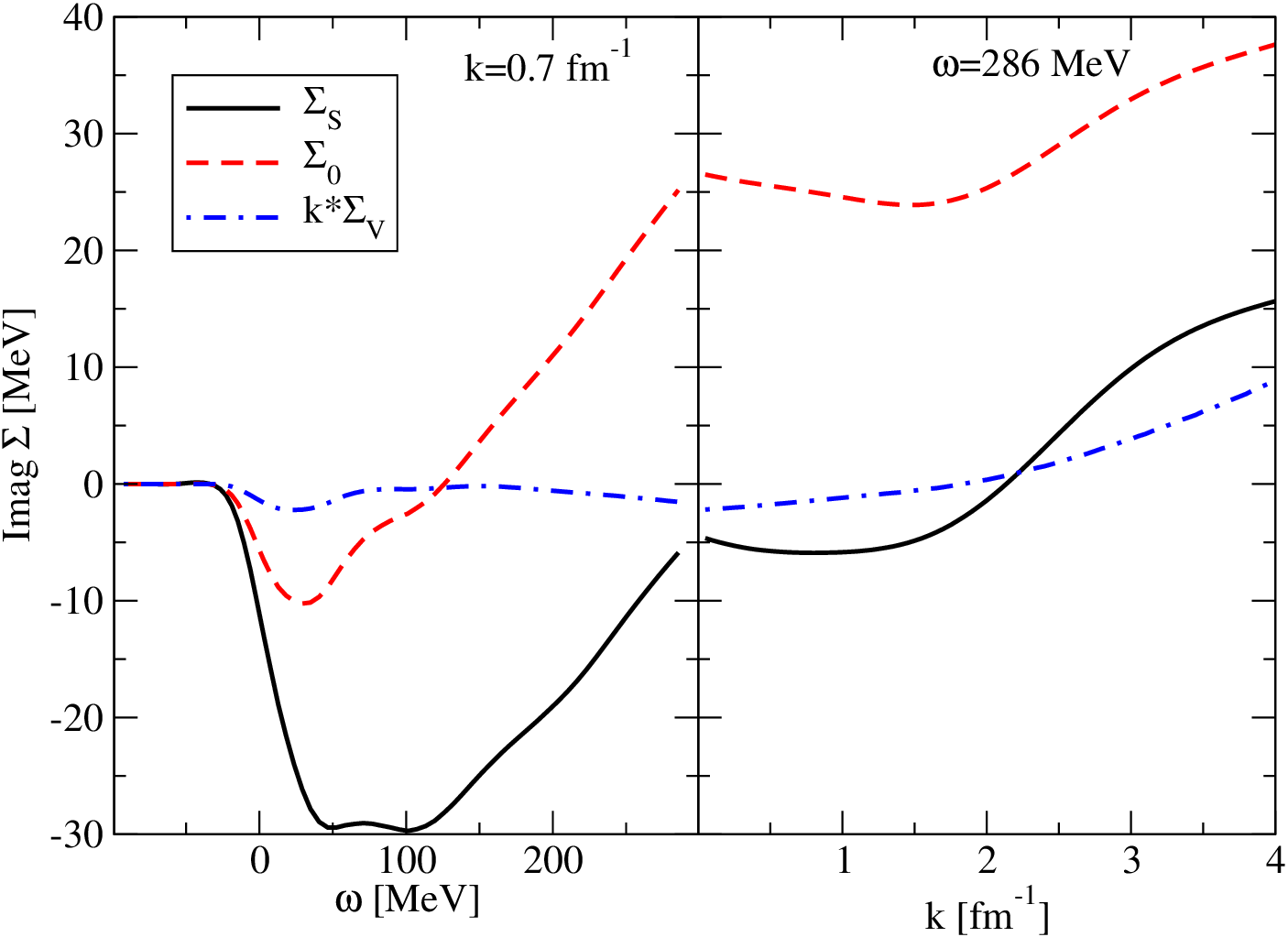}
\caption{(Color online) The imaginary part of the self-energy components calculated 
in isospin symmetric nuclear matter at a density of n$_B$=0.181 fm$^{-3}$. 
Left: the energy dependence is presented. Right: the momentum dependence is depicted.
\label{fig:imagselfenergy}}
\end{center}
\end{figure}

The energy and momentum dependence of the imaginary part of the
self-energy components at the saturation density of our EoS in isospin
symmetric nuclear matter are depicted in Fig.~\ref{fig:imagselfenergy}. 
%%%%%%%%%%%%%%%%%%%%%%%%%%%%%%%%%%%%%%%%%%%%%%%%%%%%%%%%%%%%%%%%%%%%%%%%
%%%%%%%%%%%%%%%%%%%%%%%%%%%%%%%%%%%%%%%%%%%%%%%%%%%%%%%%%%%%%%%%%%%%%%%%%
Since only particle-particle ladders are included in the solution of the BS
equation (\ref{subsec:SM;eq:BS}), which defines the $T$ matrix, these 
imaginary self-energy components are different from zero for energies
above the Fermi energy of -26.5 MeV. For energies just above this
threshold, the imaginary part of the scalar component $\Sigma_s$ as well
as of the time-like vector component $\Sigma_0$ are negative, which implies that they
tend to compensate each other in the Dirac equation for the upper
component. At larger values for the energy $\omega$ the difference
$\Sigma_s-\Sigma_0$ essentially remains constant. This is very different
from results obtained within a simple $\sigma\omega$
model\cite{trasobares:1998}, indicating that the iterated $\pi$ exchange terms
are dominating the 2 particle - 1 hole contributions to the self-energy,
when a realistic interaction model is used. The imaginary part of the
space-like vector component $\Sigma_v$ is rather small.

\begin{figure}[!ht]
\begin{center}
\includegraphics[width=0.6\textwidth] {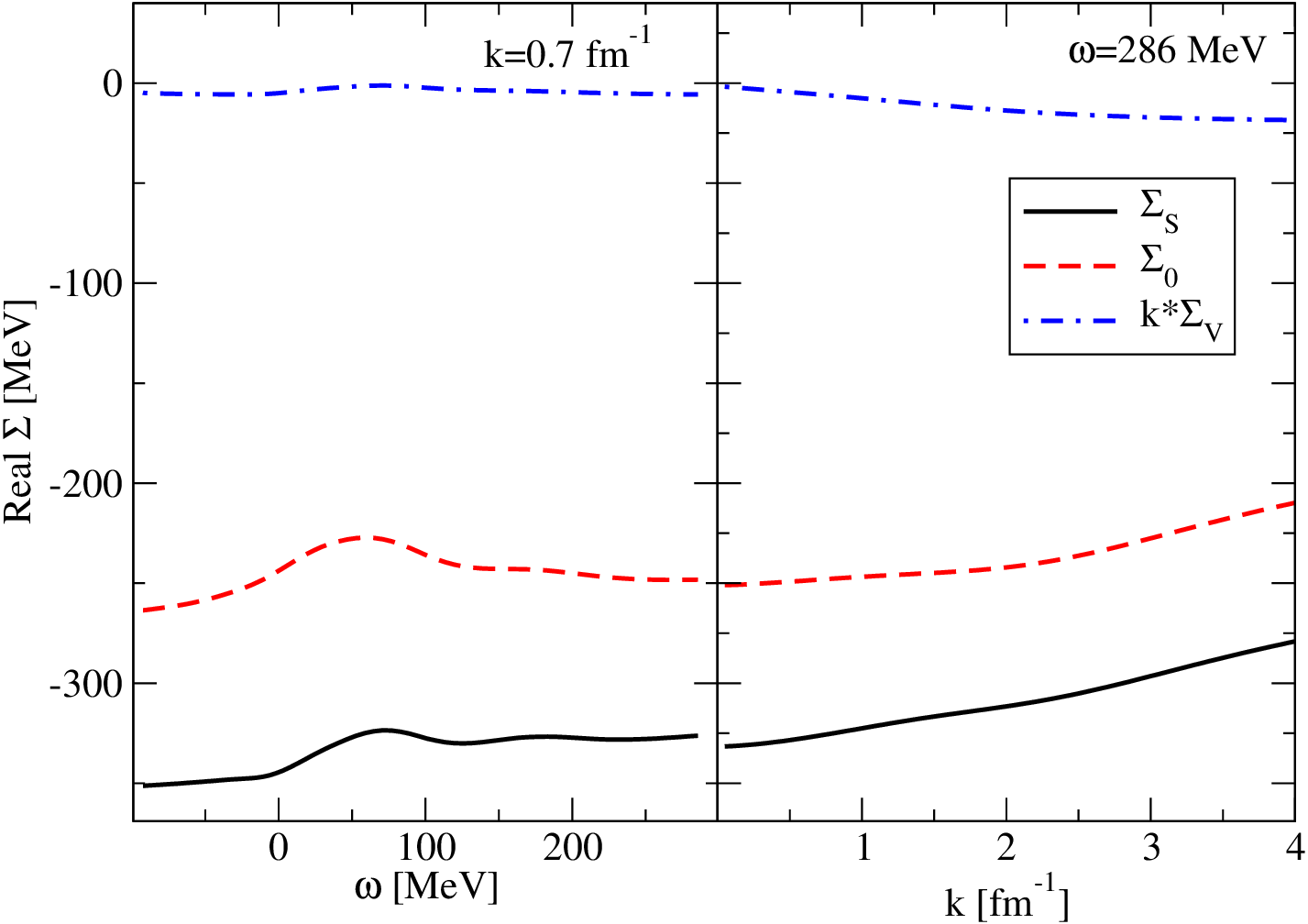}
\caption{(Color online) The real part of the self-energy components calculated in isospin symmetric nuclear matter at density of n$_B$=0.181 fm$^{-3}$. Left: the energy dependence is presented. Right: the momentum
dependence is depicted. \label{fig:realselfenergy}}
\end{center}
\end{figure}

An example for energy and momentum dependence of the real part of the
nucleon self-energy is shown in Fig.~\ref{fig:realselfenergy}. 
%%%%%%%%%%%%%%%%%%%%%%%%%%%%%%%%%%%%%%%%%%%%%%%%%%%%%%%%%%%%%%%%%%%%%%%%
%%%%%%%%%%%%%%%%%%%%%%%%%%%%%%%%%%%%%%%%%%%%%%%%%%%%%%%%%%%%%%%%%%%%%%%%%
In the energy dependence, a small enhancement appears just above the Fermi
energy of -26.5 MeV, where the imaginary self-energy components turn
nonzero.   However, the energy dependence of the real part of nucleon
self-energy is still rather weak. The moment dependence shows a very
smooth behavior.  The degree of sensitivity of the self-energy components
on energy $\omega$ and momentum k shown in Fig.~\ref{fig:realselfenergy}
is relevant for  'reference spectrum approximation' used in the iteration
procedure, since a strong momentum and energy dependence questions the
validity of the  'reference spectrum approximation'. However, this energy
and momentum dependence of the self-energy components can be characterized
as rather weak as can be seen in Fig.~\ref{fig:realselfenergy}. One must
keep in mind, however, that Fig.~\ref{fig:realselfenergy} shows two
quantities, $\Sigma_s$ and $\Sigma_0$, which are big and compensate each
other to a large extent, when inserted into the Dirac equation. Therefore
a weak dependence of these components can get magnified in solving the
Dirac equation. Therefore in the following we will use these momentum and
energy dependent components but discuss combinations of these components,
which are relevant for nuclear physics at low energies.

\subsection{Optical potentials and spectral functions}

\begin{figure}[!ht]
\begin{center}
\includegraphics[width=0.6\textwidth] {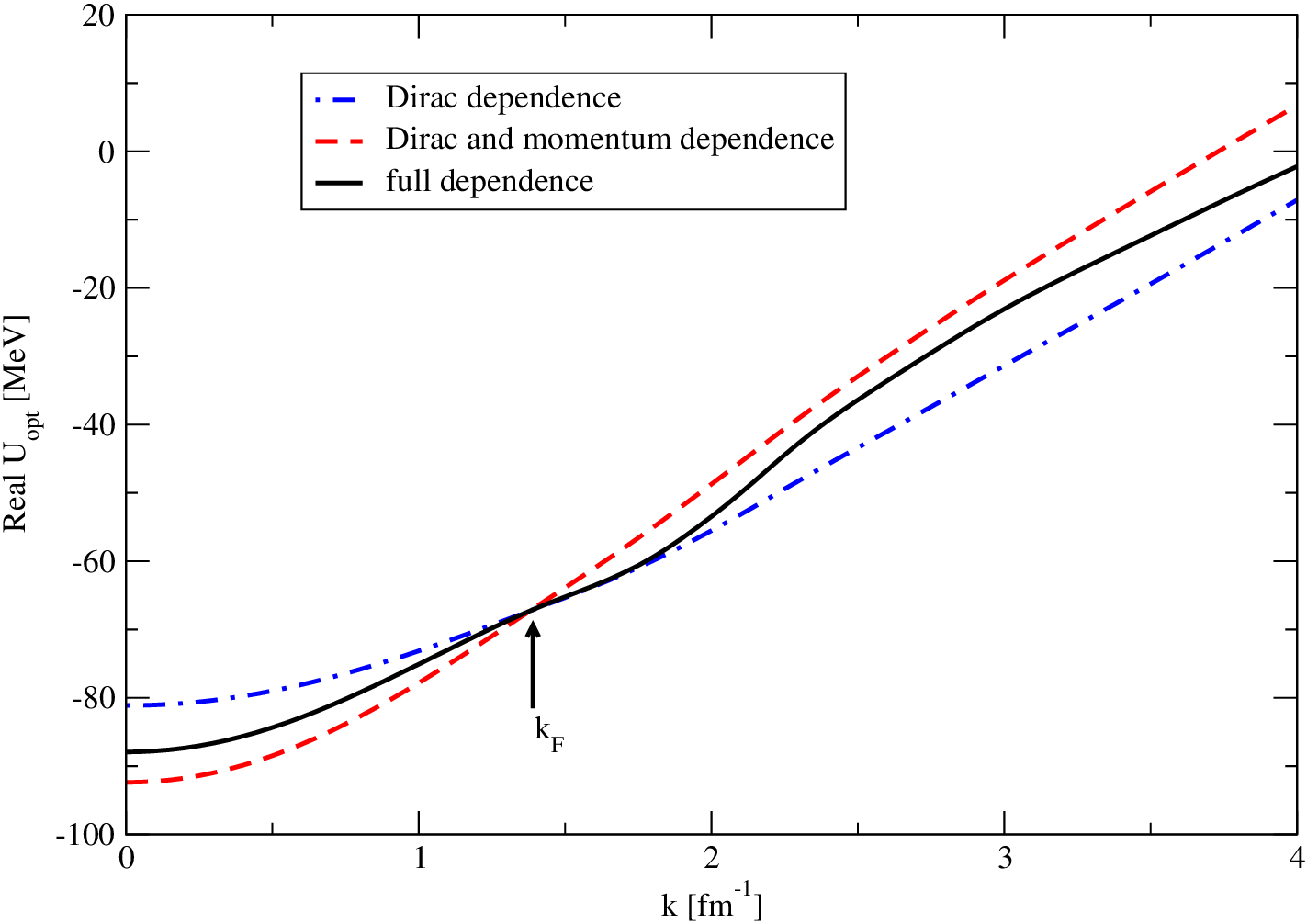}
\caption{(Color online) The real part of the on-shell optical potential as
defined in (\protect{\ref{fig:optpot}}) for $\omega =\varepsilon(|\veck|)$
for symmetric nuclear matter at  fixed nuclear density
$n_B = 0.181 \ \textrm{fm}^{-3}$. The various approximations are discussed
in the text.}
\label{fig:realopt}
\end{center}
\end{figure}

An interesting quantity is the Schr\"odinger equivalent optical potential.
This potential is obtained when the Dirac equation is reduced to an
equivalent Schroedinger equation for the large component of the Dirac
spinor. Therefore it can be identified with the non-relativistic optical
potential for a nucleon inside  the nuclear medium. This potential, 
\beqa
U(|\veck|,\omega)=\Sigs(|\veck|,\omega)- \frac{1}{M} k^{\mu} \Sigma_{\mu}(|\veck|,\omega)
+ \frac{\Sigma_s^2(|\veck|,\omega)-\Sigma_{\mu}^2(|\veck|,\omega)}{2 M},
\label{fig:optpot}
\eeqa
can be obtained from the relativistic self-energy components in
Eq.~(\ref{subsec:SM;eq:self1}). Of special interest is the on-shell value
of this optical potential, which means that we consider the case $\omega =
\varepsilon(|\veck|)$ with the single-particle energy defined below in
(\ref{epsilonk}). Results for the real part of this optical potential are
displayed in Fig.~\ref{fig:realopt} (solid line).
%%%%%%%%%%%%%%%%%%%%%%%%%%%%%%%%%%%%%%%%%%%%%%%%%%%%%%%%%%%%%%%%%%%%%%%%
%%%%%%%%%%%%%%%%%%%%%%%%%%%%%%%%%%%%%%%%%%%%%%%%%%%%%%%%%%%%%%%%%%%%%%%%%

What determines the momentum dependence of this optical potential? If one
ignores the energy and momentum dependence of relativistic self-energy
components using e.g. $\veck = k_F$ and $\omega = \varepsilon_F$ one obtains a
momentum dependence as presented by the dashed dotted line in
Fig.~\ref{fig:realopt}. This momentum dependence is a
relativistic feature as it originates from the reduction of the Dirac
equation to the non-relativistic Schroedinger equation. That is why we have
labeled this curve as the Dirac dependence. 

If in a next step the momentum dependence of the relativistic components
of the self-energy is taken into account (keeping $\omega = \varepsilon_F$)
the dashed line is obtained. We see that the inclusion of this non-locality
in space, which mainly originates from the Fock exchange term in the
self-energy tends to enhance the momentum dependence of the optical
potential (see dashed line, labeled ``Dirac and momentum dependence'').

The effects of the momentum dependence are partly compensated If also 
the energy dependence of the self-energy is considered. The full result is
rather close to the ``Dirac only'' approach in particular close to the
Fermi surface.

\begin{figure}[!ht]
\begin{center}
\includegraphics[width=0.9\textwidth] {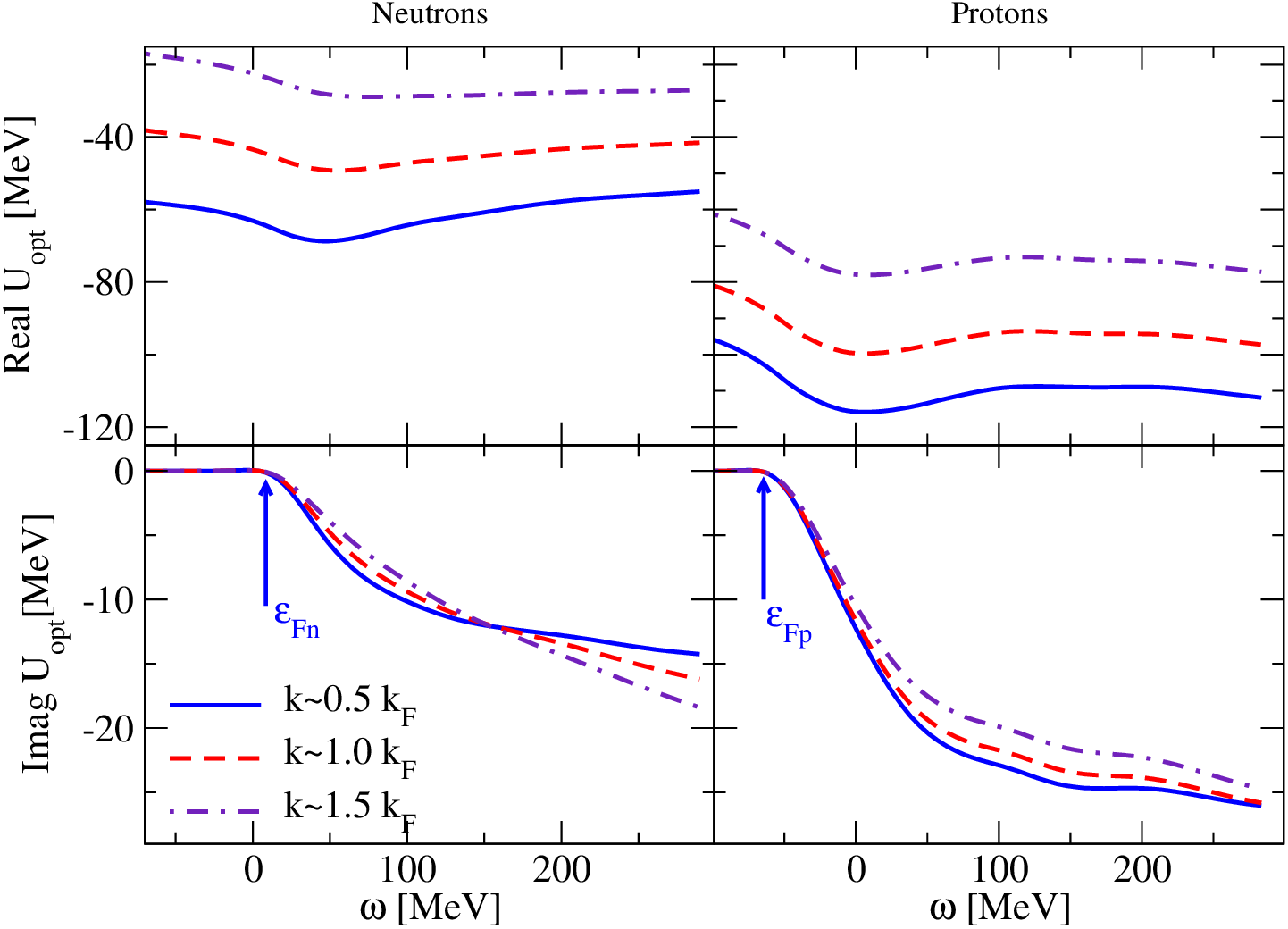}
\caption{(Color online) The energy dependence of the optical potential for
neutrons (left panels) and protons (right panels) in isospin asymmetric nuclear matter 
with an asymmetry parameter of $\beta = 0.5$ at fixed nuclear density
$n_B = 0.181 \ \textrm{fm}^{-3}$. The real part (upper panels) 
and the imaginary part (lower panels) of the neutron optical potential 
are plotted for various momenta.}
\label{fig:optpn}
\end{center}
\end{figure}

The energy dependence of the neutron and proton optical potentials in 
isospin asymmetric nuclear matter with an asymmetry parameter of $\beta =
0.5$ are plotted in Fig.~\ref{fig:optpn} for various values of the momentum k.
%%%%%%%%%%%%%%%%%%%%%%%%%%%%%%%%%%%%%%%%%%%%%%%%%%%%%%%%%%%%%%%%%%%%%%%%
%%%%%%%%%%%%%%%%%%%%%%%%%%%%%%%%%%%%%%%%%%%%%%%%%%%%%%%%%%%%%%%%%%%%%%%%%
%%%%%%%%%%%%%%%%%%%%%%%%%%%%%%%%%%%%%%%%%%%%%%%%%%%%%%%%%%%%%%%%%%%%%%%%
%\begin{figure}[!h]
%\begin{center}
%\includegraphics[width=0.9\textwidth] {optpotproton.eps}
%\caption{(Color online) The energy dependence of the proton optical potential in isospin asymmetric nuclear matter 
%with an asymmetry parameter of $\beta = 0.5$ at fixed nuclear density
%$n_B = 0.181 \ \textrm{fm}^{-3}$. The real part (upper panel) and the imaginary part (lower panel) of the proton optical potential are plotted.}
%\label{fig:optpotproton}
%\end{center}
%\end{figure}
%%%%%%%%%%%%%%%%%%%%%%%%%%%%%%%%%%%%%%%%%%%%%%%%%%%%%%%%%%%%%%%%%%%%%%%%%
The lower panels show the corresponding imaginary parts of these
potentials. These imaginary parts are identical to zero for energies
$\omega$ less than the corresponding Fermi energy, i.e. $\omega<\varepsilon_F$. 
At energies just above the Fermi energy, they initially decrease with a   a steep
negative slope and then seem to stabilize. This stabilization is identical
to the example of symmetric nuclear matter, as we discussed before in
connection with Fig.~\ref{fig:imagselfenergy}. It should be recalled,
however, that at smaller energy the main contribution originates from the
imaginary part of $\Sigma_s$, whereas at energies $\omega >$  200 MeV the
vector component $\Sigma_0$ tends to dominate. The momentum dependence of
the imaginary part is rather weak.

The real part of the optical potential gets more attractive with
increasing energy until one reaches values of the energy at which the
imaginary part  is different from zero. The real part then turns less
attractive at higher energies. Therefore, the energy dependence of the
real part of the optical potential displays a minimum at energies just
above the Fermi energies  as can be seen in the upper panels of
Fig.~\ref{fig:optpn}.  Such a minimum
around the Fermi energy is also found in the  self-energy from
non-relativistic BHF calculations~\cite{frick:2002}. Another observation
made from Fig.~\ref{fig:optpn} 
concerns the momentum dependence. It is found that the real part of
optical potential becomes less attractive with increasing momenta.
 
In isospin asymmetric nuclear matter, the properties of neutrons and 
protons differ from each other as one can see comparing the panels on the
left and right side of Fig.~\ref{fig:optpn}. 
The real part of the proton optical potential is
more attractive than that of the neutron optical potential in neutron-rich
matter. Also the absolute values for the imaginary part are larger for protons
than for the neutrons. These results are easy to understand from the fact
that the proton neutron interaction is stronger than the neutron neutron
or proton proton interactions. Therefore the protons are exposed to a
stronger mean field which is caused mainly from the interaction with the
large number of neutrons around.

\begin{figure}[!t]
\begin{center}
\includegraphics[width=0.6\textwidth] {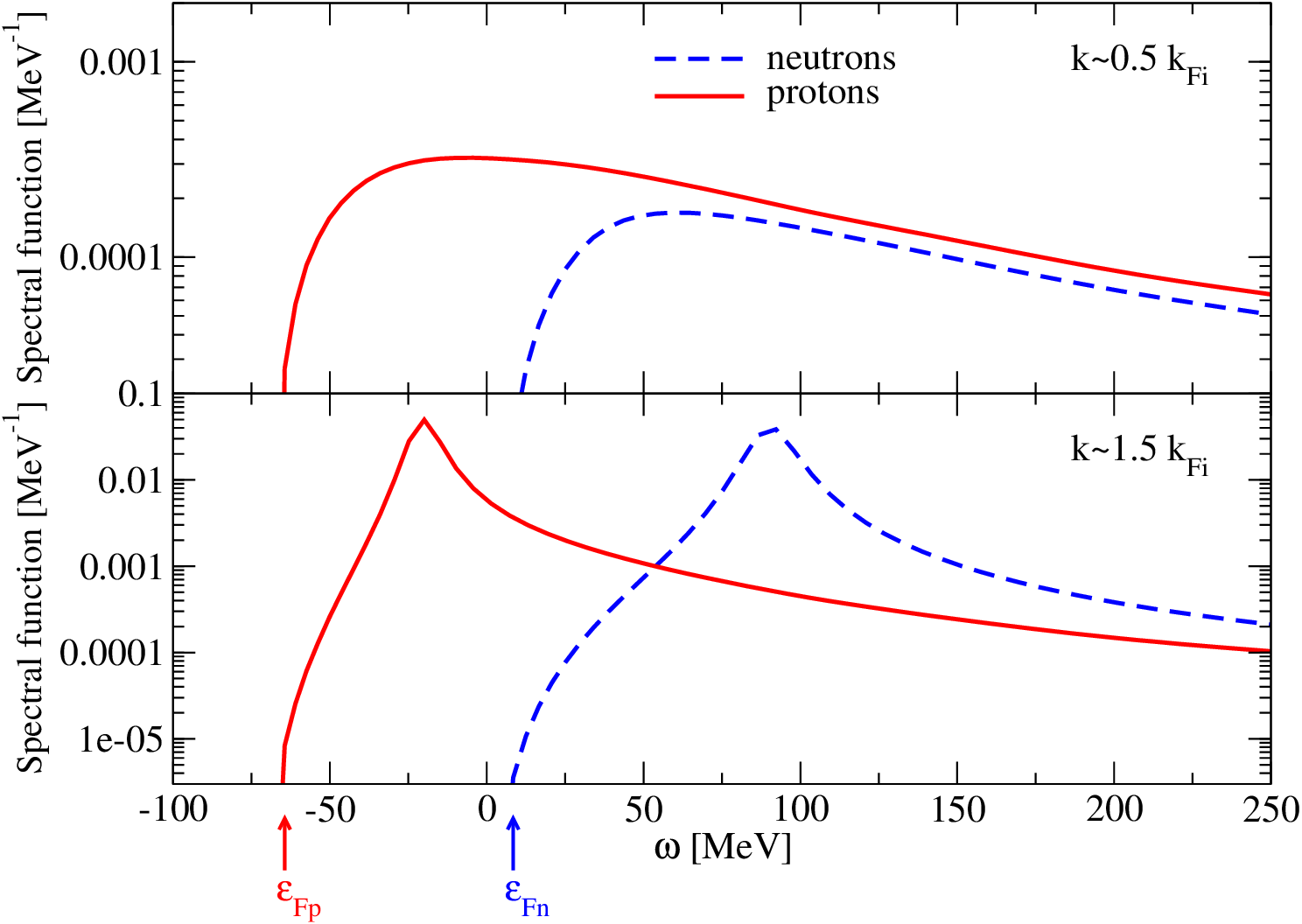}
\caption{(Color online) Particle spectral functions for nucleons with  $k \sim$
0.5 $k_{Fi}$ in the upper part and $k \sim$1.5 $k_{Fi}$ in the lower panel as a 
function of energy $\omega$ in isospin asymmetric nuclear matter  with an
asymmetry parameter of $\beta =0.5$ at fixed nuclear density of  $n_B = 0.181 \
\textrm{fm}^{-3}$.}
\label{fig:spectral}
\end{center}
\end{figure}

This real and imaginary part of the optical potential can also be used to
determine the  spectral function for the particle strength from its
non-relativistic definition,
\beqa
S^p(|\veck|,\omega)=-\frac{1}{\pi} \frac{\Im m
U(|\veck|,\omega)}{[\omega-k^2/2M-\Re e U(|\veck|,\omega)]^2+[\Im m
U(|\veck|,\omega)]^2},
\eeqa
for $\omega>\varepsilon_F$.
It represents the probability that a nucleon with momentum k and energy $\omega$
can be added to the ground state. Fig.~\ref{fig:spectral} displays the spectral
functions for protons and neutrons in isospin asymmetric nuclear matter  with an
asymmetry parameter of $\beta =0.5$ at fixed nuclear density of $n_B = 0.181 \
\textrm{fm}^{-3}$.
%%%%%%%%%%%%%%%%%%%%%%%%%%%%%%%%%%%%%%%%%%%%%%%%%%%%%%%%%%%%%%%%%%%%%%%%
%%%%%%%%%%%%%%%%%%%%%%%%%%%%%%%%%%%%%%%%%%%%%%%%%%%%%%%%%%%%%%%%%%%%%%%%%

The upper part of this figure shows the particle strength for momenta below the
corresponding Fermi momenta for protons and neutrons. In the independent
particle model states with these momenta would be completely occupied and the
particle strength is identical to zero. Since, however, the Brueckner G-matrix
accounts for particle-particle ladders, the BHF and also the DBHF self-energies
include the effects of 2 particle - 1 hole terms, which lead to a non-vanishing
imaginary part for $\omega>\varepsilon_F$. Due to these 2 particle - 1 hole
components we observe a non-vanishing spectral particle strength for momenta
below $k_F$. From the upper part of Fig.~\ref{fig:spectral} we can see that the 
larger values of the  imaginary part of the proton optical
potential displayed in Fig.~\ref{fig:optpn} lead to larger values for the
proton spectral functions than for the neutron spectral functions. This has also
been observed in non-relativistic calculations of asymmetric nuclear 
matter~\cite{hassaneen:2004}.

This non-vanishing particle strength for momenta below $k_F$ should be
accompanied by a depletion of the occupation number below 1 for these states.
Note, however, that the BHF approach as well as the DBHF approximation is not
number conserving. As it does not account for hole-hole ladder terms one does
not obtain a spectral distribution for energies $\omega <\varepsilon_F$. The
depletion of the occupation numbers for the hole states ($k < k_F$), however, 
can be determined from the single-particle strength at the quasi particle poles
of the single-particle Greens function~\cite{wimrep:1992} 
\beqa
z(\veck) = \left\{1-\left(\frac{\partial \Re e U(|\veck|,\omega)}{\partial
\omega}\right)_{\omega=\varepsilon(k)}\right\}^{-1}\,.
\eeqa
Since the energy dependence of the real part of the optical potential in
neutron rich matter is larger for the protons than for the neutrons (see
Fig.~\ref{fig:optpn}) we obtain larger depletions for the protons than for the
neutrons. While the neutron occupation number varies between 0.95 for $k\approx 0.5
k_{Fn}$ and 0.87 for $k\approx k_{Fn}$, the corresponding numbers for the proton are
0.87 ($k\approx 0.5 k_{Fp}$) and 0.8 ($k\approx k_{Fp}$). The stronger proton neutron
interaction yields a larger depletion for the protons than for the neutrons in
neutron rich matter.

The lower panel of Fig.~\ref{fig:spectral} exhibits the particle strength
distribution for momenta larger than the Fermi momentum. The imaginary part of
the self-energy leads to a broad distribution of the single-particle strength.

\subsection{Single-particle energy}
\label{subsec:spenergy}

\begin{figure}[!t]
\begin{center}
\includegraphics[width=0.7\textwidth] {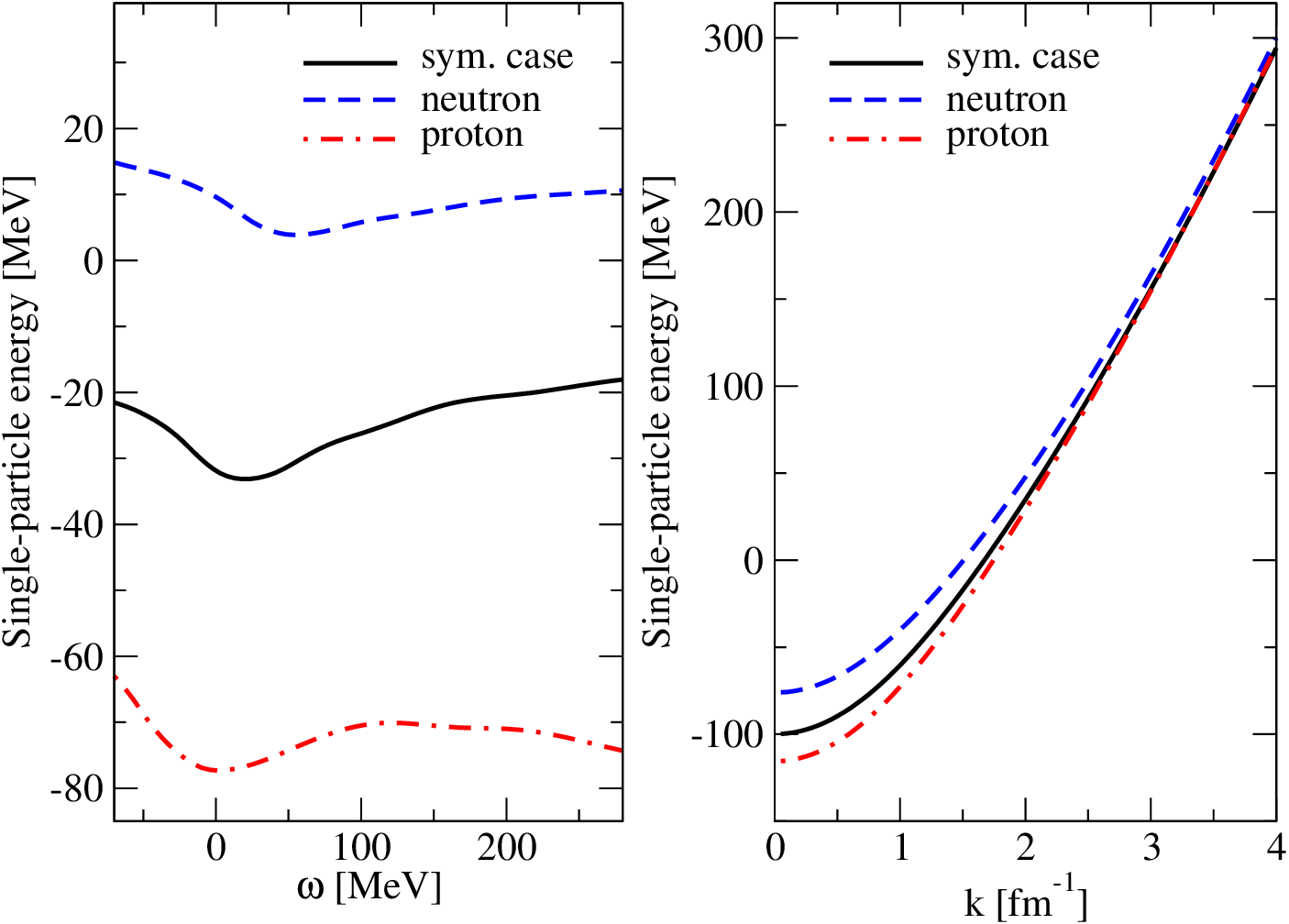}
\caption{(Color online) Energy and momentum dependence of the single-particle energy. The neutron (dashed line) and proton (dashed-dotted line) 
single-particle energies are depicted for isospin asymmetric nuclear matter with an asymmetry parameter of $\beta =0.5$ at a fixed nuclear density of $n_B = 0.181 \ \textrm{fm}^{-3}$.
The nucleon single-particle energy in isospin symmetric nuclear matter (solid
line) is also given . Left: energy dependence at $k=k_{Fi}$. Right: momentum
dependence at $\omega=\varepsilon_{Fi}$.}
\label{fig:kwdepspenergy}
\end{center}
\end{figure}

The relativistic expression of the single-particle energy is given by
\beqa
\varepsilon(|\veck|,\omega)=- \Sigo (|\veck|,\omega) + \left(1.+\Sigv
(|\veck|,\omega)\right) \sqrt{\veck^2+
\left(\frac{M+\Sigs(|\veck|,\omega)}{1.+ \Sigv (|\veck|,\omega)}\right)^2
}-M.\label{epsilonk}
\eeqa
Energy and momentum dependence of the single-particle energy in isospin symmetric and asymmetric nuclear are plotted in Fig.~\ref{fig:kwdepspenergy}.
%%%%%%%%%%%%%%%%%%%%%%%%%%%%%%%%%%%%%%%%%%%%%%%%%%%%%%%%%%%%%%%%%%%%%%%%
%%%%%%%%%%%%%%%%%%%%%%%%%%%%%%%%%%%%%%%%%%%%%%%%%%%%%%%%%%%%%%%%%%%%%%%%%
The energy dependence of the single-particle potential in the left panel displays a minimum at energies just above the Fermi energies, which is related to the small enhancement in the real part of the self-energy. In the right panel, a rough quadratic dependence of the single-particle energy on the momentum k is found. Such a quadratic dependence is often assumed in non-relativistic calculations~\cite{frick:2002},
\beqa
\varepsilon \approx \frac{\veck^2}{2 M^*}+C.\label{eq:epsilonofk}
\eeqa
Furthermore, in Fig.~\ref{fig:kwdepspenergy}  the neutron has a higher single-particle energy than the proton due to its less attractive potential in neutron-rich matter.

\subsection{Effective Mass}
A common concept in the field of nuclear physics is the effective mass. 
However, the expression of an effective nucleon mass  has been used in various connections in
many-body physics and to denote
different quantities. This includes the non-relativistic effective mass $m^*_{NR}$ and the
relativistic Dirac mass $m^*_{D}$. 

\begin{figure}[!t]
\begin{center}
\includegraphics[width=0.7\textwidth] {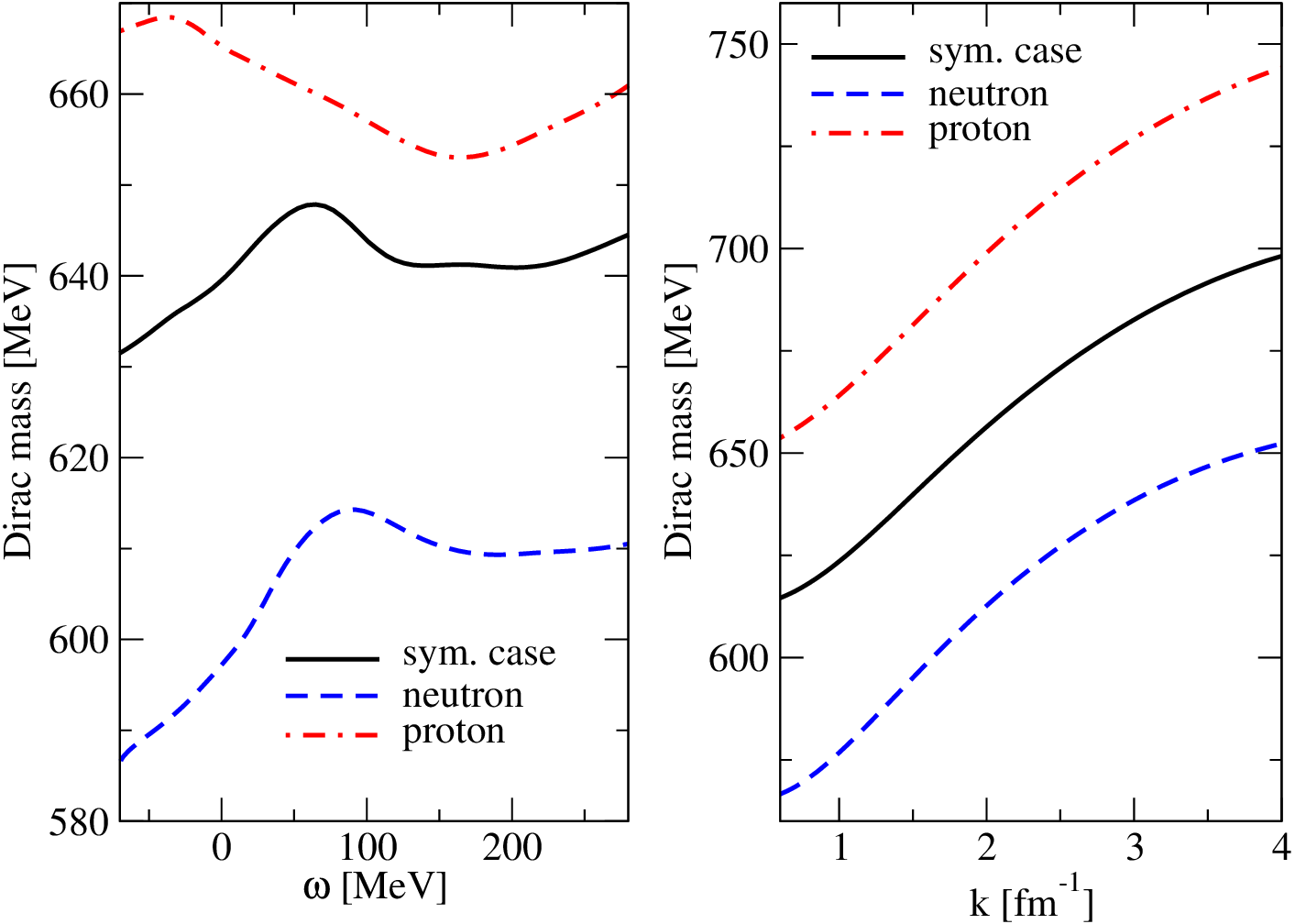}
\caption{(Color online) Energy and momentum dependence of the Dirac mass. The neutron (dashed line) and proton (dashed-dotted line) Dirac masses are depicted for isospin asymmetric nuclear matter with asymmetry parameter $\beta =0.5$ at a fixed nuclear density of $n_B = 0.181 \ \textrm{fm}^{-3}$.
The nucleon Dirac mass in isospin symmetric nuclear matter (solid line) is also
given. Left: energy dependence at $k=k_{Fi}$. Right: momentum dependence at
$\omega=\varepsilon_{Fi}$.}
\label{fig:kwdepDiracmass}
\end{center}
\end{figure}

The Dirac mass is a genuine relativistic quantity and can only be
obtained from relativistic many-body approaches.
The effective Dirac mass accounts for medium effects through the scalar part 
of the self-energy. It is given by
\beqa
m^*_D(|\veck|,\omega) =\frac{M + \Re e \Sigma_s(|\veck|, \omega)}{1+\Re e \Sigv(|\veck|,\omega)}.
\eeqa
The energy and momentum dependency of this Dirac mass are plotted in Fig.~\ref{fig:kwdepDiracmass}. 
%%%%%%%%%%%%%%%%%%%%%%%%%%%%%%%%%%%%%%%%%%%%%%%%%%%%%%%%%%%%%%%%%%%%%%%%
%%%%%%%%%%%%%%%%%%%%%%%%%%%%%%%%%%%%%%%%%%%%%%%%%%%%%%%%%%%%%%%%%%%%%%%%%
The maximum in the Dirac mass just above the Fermi energy in the left panel in Fig.~\ref{fig:kwdepDiracmass} originates 
from the small enhancement in the scalar self-energy. In the right panel,
the smooth behavior of the momentum dependence can be observed. In addition, it can be observed that the effective Dirac mass of the proton is larger
than that of the neutron. This result of the larger proton Dirac mass in neutron-rich matter has been mentioned in previous works of DBHF calculations based on projection techniques~\cite{dejong:1998,schiller:2001,vandalen:2004b,vandalen:2005a,vandalen:2005b,vandalen:2007}. 

In contrast, the non-relativistic mass is the result of a quadratic
parameterization of the single-particle spectrum mentioned in the
section~\ref{subsec:spenergy} (see eq.(\ref{eq:epsilonofk})). It is a measure of
the non-locality of the single-particle potential $U$.  Therefore, the effective
non-relativistic mass is given by
\begin{equation}
m^*_{NR}(|\veck|,\omega=\varepsilon(|\veck|,\omega)) = \left[ \frac{1}{M} + \frac{1}{|\veck|}
    \frac{\partial U (|\veck|,\omega=\varepsilon(|\veck|,\omega)) }{\partial |\veck|} \right]^{-1}.
\end{equation}
The non-locality of $U$ can be due to non-localities in space, which results in a momentum dependence or in time, which results in an energy dependence. In order to separate both effects, these two types of non-localities have been characterized by the $k$-mass,
\begin{equation}
m^*_k(|\veck|,\omega) = \left[ \frac{1}{M} + \frac{1}{|\veck|}
    \frac{\partial U (|\veck|,\omega) }{\partial |\veck|} \right]^{-1},
\end{equation},
 and by the $E$-mass,
\begin{equation}
m^*_E(|\veck|,\omega) = M\left[1 - 
    \frac{\partial U (|\veck|,\omega) }{\partial \omega} \right] ,
\end{equation}
respectively. 
These masses can be determined from both, as well relativistic as non-relativistic approaches. 

\begin{figure}[!t]
\begin{center}
\includegraphics[width=0.7\textwidth] {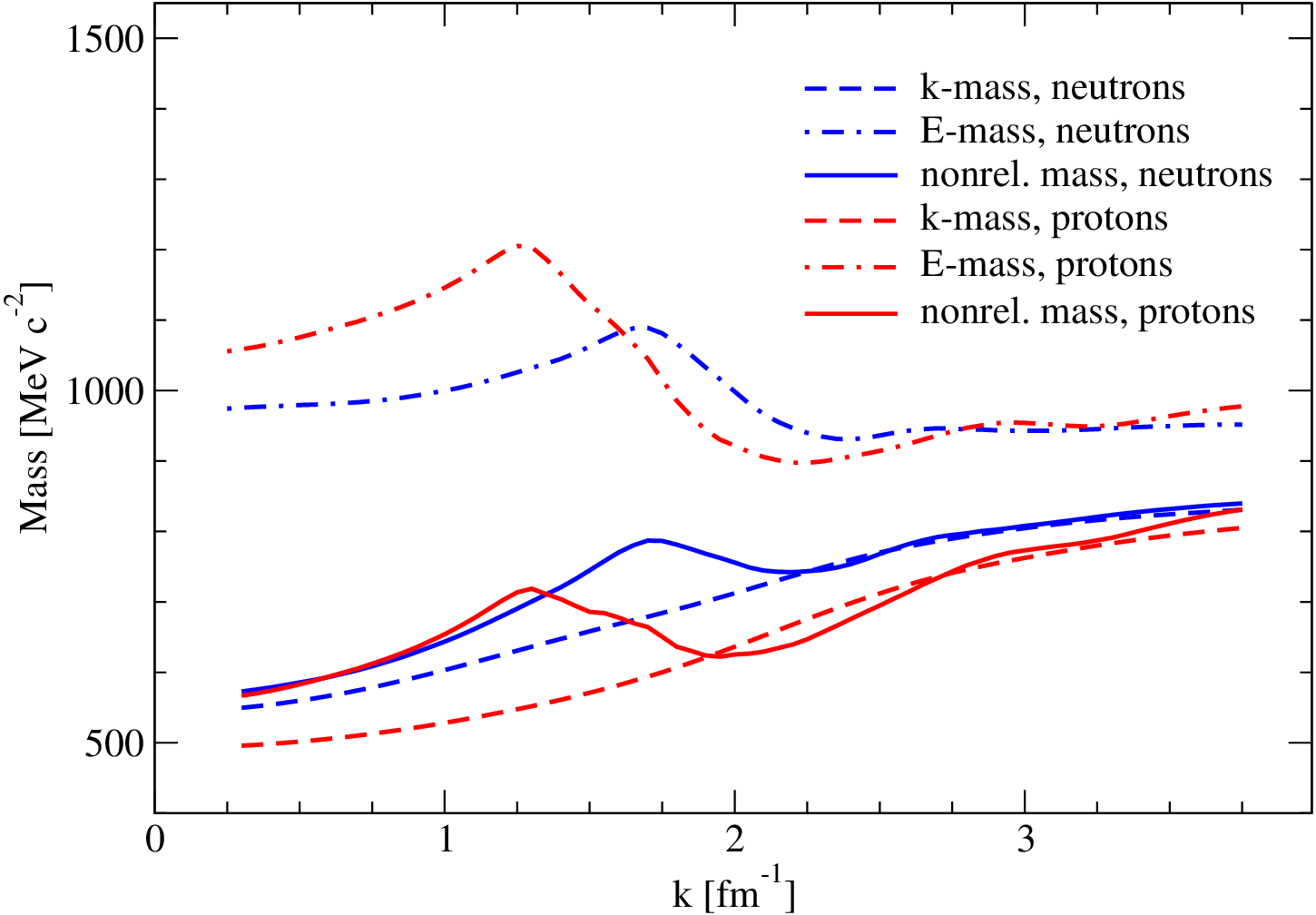}
\caption{(Color online) The effective non-relativistic mass (solid lines), the effective $k$-mass (dashed lines), and
the effective $E$-mass (dashed-dotted lines) at the on-shell point, i.e.
$\omega=\varepsilon(|\veck|,\omega)$, for neutrons and protons as obtained 
from relativistic DBHF calculations for isospin asymmetric nuclear matter
at a density of $\rho=0.181 \text{ fm}^{-3}$ and a proton abundance of 25 \% ($\beta$=0.5).}
\label{fig:Ekmass}
\end{center}
\end{figure}

In Fig.~\ref{fig:Ekmass}, the presented masses at the on-shell point, i.e.
$\omega=\varepsilon(|\veck|,\omega)$, are obtained from our relativistic DBHF calculation  using Eq.~(\ref{fig:optpot}). 
%%%%%%%%%%%%%%%%%%%%%%%%%%%%%%%%%%%%%%%%%%%%%%%%%%%%%%%%%%%%%%%%%%%%%%%%
%%%%%%%%%%%%%%%%%%%%%%%%%%%%%%%%%%%%%%%%%%%%%%%%%%%%%%%%%%%%%%%%%%%%%%%%%
The pronounced peak of the non-relativistic mass slightly above $\kf$ as is also seen in non-relativistic 
Green's function calculations~\cite{ramos:1990} and BHF calculations~\cite{jaminon:1989,hassaneen:2004,gogelein:2009} is reproduced. This peak structure of the non-relativistic mass is the result 
of subtle cancellation effects of the scalar and vector self-energy components in the relativistic framework. 
Therefore, a very precise method is required in order to determine variations of the 
self-energy, since they are small compared to their absolute scale. The 
applied projection techniques are the adequate tool for this purpose, whereas 
the extraction of mean self-energy components from a fit to 
the single-particle potential~\cite{alonso:2003} is not able to resolve such 
a structure at all. 

Another issue concerns isospin asymmetric properties, i.e. the proton-neutron mass 
splitting. Although the Dirac mass derived from the DBHF approach has a proton-neutron mass 
splitting of $m^*_{D,n} <m^*_{D,p}$ as can be seen from Fig.~\ref{fig:kwdepDiracmass}, 
the non-relativistic mass derived from the DBHF approach 
shows the opposite behavior, i.e. $m^*_{NR,n} > m^*_{NR,p}$, 
which is in agreement with the results from non-relativistic 
BHF calculations~\cite{zuo:1999,hassaneen:2004}. This has been investigated earlier in the works of Refs.~\cite{vandalen:2005a,vandalen:2005b}.
However, the $k$-mass and $E$-mass from these relativistic approaches are not considered in these works, since the determination of these two masses requires the knowledge of the off-shell behavior of the single-particle potential $U$. 

These $k$-masses and $E$-masses obtained from our relativistic DBHF calculations are plotted in Fig.~\ref{fig:Ekmass} for isospin asymmetric nuclear matter at a density of $\rho$ = 0.181 fm$^{-3}$ and an asymmetry parameter of $\beta$=0.5. The effective $k$-mass, which corresponds to the non-localities in space of single-particle potential, are mainly 
generated by exchange Fock terms.  It can be observed that the resulting $k$-mass is a smooth 
function of the momentum, which is also in agreement with results from non-relativistic calculations~\cite{frick:2002}.
Another observation is that the effective $k$-mass for the protons 
is significantly below the corresponding value for the neutrons at all momenta. This result also is in agreement with results obtained from non-relativistic BHF calculations~\cite{hassaneen:2004,gogelein:2009}. 

The effective $E$-mass represents the non-locality in time. This non-locality in time is generated by Brueckner 
ladder correlations due to the scattering to intermediate states which are off-shell. These are mainly short-range correlations which generate a 
strong momentum  dependence with a characteristic enhancement of the 
$E$-mass slightly above the Fermi surface as can be observed in Fig.~\ref{fig:Ekmass}. The maximum value is even higher
than the bare mass $M$. This peak structure is also observed 
in the case of non-relativistic calculations~\cite{mahaux:1985,jaminon:1989,frick:2002,hassaneen:2004,gogelein:2009}. Therefore, the enhancement of the non-relativistic mass is due to the
effective $E$-mass. Since the effective $E$-mass is not strong enough to compensate for
the effects of the $k$-mass, the effective non-relativistic mass for neutrons remains larger than the
corresponding one for protons.

\section{Summary and Conclusion}
\label{sec:S&C}
In this work we describe the off-shell behavior of nucleon properties in isospin
asymmetric nuclear  matter in the relativistic DBHF approach based on projection
techniques using the Bonn A potential. In addition, the optimal representation
scheme for the $T$-matrix, the subtracted $T$-matrix representation, is applied.
At the end of the iteration procedure, we keep not only the momentum dependence
but also the explicit energy dependence of the relativistic components of the
self-energy for our investigation of  the off-shell
behavior of nucleon properties in isospin asymmetric nuclear matter. These
off-shell effects are relevant for reactions occurring in radioactive beam
experiments.

An issue considered is the off-shell behavior of the optical potential and the
related spectral function. Since the BHF approximation does not account for
hole-hole ladder terms the imaginary part of the relativistic
self-energy components are identical to zero for energies below the Fermi
energy. As a consequence also the imaginary part of the optical potential and spectral function are
identical to zero in this energy range. However, these quantities yield
non-negligible values above the Fermi energy. The real part of the optical
potential yields nonzero values in the entire energy range considered and
displays a minimum at energies just above the Fermi energies. Furthermore, the
real and the imaginary part of the proton optical potential are much stronger
than those of the neutron optical potential in neutron-rich matter. This is due
to the stronger proton-neutron as compared to the neutron-neutron and
proton-proton interactions. These larger
values of the  imaginary part of the proton optical potential also lead to
larger values for the particle spectral functions of hole states and the
corresponding depletions of the occupation numbers for the hole states. 
This behavior has also be observed in
non-relativistic BHF calculations~\cite{hassaneen:2004}.

Another issue is the behavior of the non-relativistic mass, which can be
determined from as well relativistic as non-relativistic approaches. The 
pronounced peak of the on-shell non-relativistic mass slightly above $\kf$, which
is typical for non-relativistic
calculations~\cite{jaminon:1989,hassaneen:2004,gogelein:2009}, is reproduced in
our relativistic calculation. This non-relativistic mass is a measure of the
non-locality in space and in time. Non-localities in space, which result in a
momentum dependence, are characterized by the $k$-mass, whereas non-localities in
time, which result in an energy dependence, are characterized by the $E$-mass.
Therefore, even the determination of the on-shell values of these quantities
require the knowledge of the off-shell behavior of the single-particle
potential. The effective $k$-mass shows a smooth behavior, whereas the $E$-mass
exhibits a large peak slightly above the Fermi surface. Therefore, the observed
strong enhancement of the non-relativistic mass is due to the behavior of the
$E$-mass. These predictions of the $k$- and $E$-mass are in agreement with
results from non-relativistic calculations~\cite{frick:2002}.

An observation concerning the isospin effects of these quantities is that the
effective $k$-mass for the protons is significantly below the corresponding
value for the neutrons. Due to the fact that the effective $E$-mass is not
strong enough to compensate for the effects of the $k$-mass, the effective
non-relativistic mass for neutrons remains larger than the corresponding one for
protons. This result for the non-relativistic mass splitting, which is opposite
to the Dirac mass splitting of $m^*_{D,n}
<m^*_{D,p}$~\cite{vandalen:2005a,vandalen:2005b}, is in agreement with the
results from non-relativistic BHF calculations~\cite{zuo:1999,hassaneen:2004}. 

Therefore, in the framework of the relativistic DBHF approach we are able to
obtain results for the off-shell behavior of nucleon properties  in as well
isospin symmetric as isospin asymmetric nuclear matter. These results for the
nucleon properties such as nucleon optical potentials, spectral functions,
single-particle energies, and effective masses, can be applied in the
description of nucleon-nucleon collisions occurring in radioactive beam
experiments.

%%%%%%%%%%%%%%%%%%%%%%%%%%%%%%%%%%%%%%%%%%%%%%%%%%%%%%%%%%%%%%%%%%%%%%%%
\begin{acknowledgments}
This work has been supported by
the Deutsche Forschungsgemeinschaft (DFG) under contract no. Mu 705/5-2.
\end{acknowledgments}
%%%%%%%%%%%%%%%%%%%%%%%%%%%%%%%%%%%%%%%%%%%%%%%%%%%%%%%%%%%%%%%%%%%%%%%%

%%%%%%%%%%%%%%%%%%%%%%%%%%%%%%%%%%%%%%%%%%%%%%%%%%%%%%%%%%%%%%%%%%%%%%%%

%%%%%%%%%%%%%%%%%%%%%%%%%%%%%%%%%%%%%%%%%%%%%%%%%%%%%%%%%%%%%%%%%%%%%%%%


\begin{thebibliography}{99}
%
\bibitem{bethe:1990} H. A. Bethe, Rev. Mod. Phys. \textbf{62}, 801 (1990). 
%
\bibitem{pethick:1995} C. J. Pethick, D. G. Ravenhall, and C. P. Lorentz, Nucl. Phys. \textbf{A584}, 675
(1995). 
%
\bibitem{vandalen:2003} E. N. E. van Dalen, A. E. L. Dieperink and J. A. Tjon, Phys. Rev. C \textbf{67}, 065807 (2003).
%
\bibitem{gogelein:2008} P. G\"{o}gelein, E. N. E. van Dalen, C. Fuchs, and H. M\"{u}ther, Phys. Rev C \textbf{77}, 025802 (2008).
%
\bibitem{tanihata:1995} I. Tanihata, Prog. Part. Nucl. Phys. \textbf {35}, 505  (1995).
%
\bibitem{hansen:1995} P. G. Hansen, A. S. Jensen, and B. Jonson, Annu. Rev. Nucl. Part. Sci. \textbf{45}, 591 (1995).
%
\bibitem{anastasio:1983} M. R. Anastasio, L. S. Celenza, W. S. Pong, and C. M. Shakin, Phys. Rep. \textbf{100}, 327 (1983).
%
\bibitem{horowitz:1987} C. J. Horowitz and B. D. Serot, Nucl. Phys. \textbf{A464}, 613 (1987).
%
\bibitem{brockmann:1990} R. Brockmann and R. Machleidt, Phys. Rev. C \textbf{42},  1965 (1990).
%
\bibitem{brown:1987} G. E. Brown, W. Weise, G. Baym and J. Speth,
Comments Nucl. Part. Phys. \textbf{17}, 39 (1987).
%
\bibitem{mahaux:1980} M. Jaminon, C. Mahaux and P. Rochus, Phys. Rev. C
\textbf{22}, 2027 (1980).
%
\bibitem{kleinm:1994} M. Kleinmann, R. Fritz, H. M\"uther and A. Ramos,
Nucl. Phys. \textbf{A579}, 85 (1994).
%
\bibitem{ulrych:1997} S. Ulrych and H. M\"uther, Phys. Rev. C \textbf{56},
1788 (1997).
%
\bibitem{dejong:1998} F. de Jong and H. Lenske, Phys. Rev. C \textbf{58}, 890  (1998).
%
\bibitem{alonso:2003} D. Alonso and F. Sammarruca, Phys. Rev. C \textbf{67}, 054301 (2003).
%
\bibitem{vandalen:2004b}
E. N. E. van Dalen, C. Fuchs, and A. Faessler, Nucl. Phys. \textbf{A744}, 227 (2004).
%
\bibitem{vandalen:2007}
E. N. E. van Dalen, C. Fuchs, and A. Faessler, Eur.Phys.J. A \textbf{31}, 29 (2007).
%
\bibitem{zuo:1999} W. Zuo, I. Bombaci, and U. Lombardo, Phys. Rev. C \textbf{60},  024605 (1999).
%
\bibitem{hassaneen:2004} Kh.S.A. Hassaneen and H. M\"uther, Phys. Rev. C \textbf{70}, 054308 (2004).
%
\bibitem{wimrep:1992} W.H. Dickhoff and H. M\"uther, Rep. on Prog. in Phys.
\textbf{11}, 1947 (1992).
%
\bibitem{dejong:1991} F. de Jong and R. Malfliet, Phys. Rev C \textbf{44}, 998 (1991).
%
\bibitem{dejong:1996} F. de Jong and H. Lenske, Phys. Rev. C \textbf{54}, 1488 (1996).
%
\bibitem{machleidt:1989} R. Machleidt, Adv. Nucl. Phys. \textbf{19}, 189 (1989).
%
\bibitem{bethe:1963}
H. A. Bethe, B. H. Brandow, and A. G. Petschek, Phys. Rev. \textbf{129}, 225 (1963).
%
\bibitem{sehn:1997} L. Sehn, C. Fuchs, and A. Faessler, Phys. Rev. C \textbf{56}, 216 (1997).
%
\bibitem{tjon:1985a} J. A. Tjon and S. J. Wallace, Phys. Rev. C \textbf{32}, 267 (1985).
%
\bibitem{trasobares:1998} A. Trasobares, A. Polls, A. Ramos and H.
M\"uther, Nucl. Phys. \textbf{A640}, 471 (1998)
%
\bibitem{frick:2002} T. Frick, Kh. Gad, H. M\"uther, and P. Czerski, Phys. Rev. C \textbf{65},  034321 (2002)
%
\bibitem{schiller:2001} E. Schiller and H. M$\ddot{\textrm{u}}$ther, Eur. Phys. J. A \textbf{11}, 15 (2001). 
%
\bibitem{vandalen:2005a} E.N.E. van Dalen, C. Fuchs, and  A. Faessler, Phys. Rev. Lett. \textbf{95}, 022302 (2005).
%
\bibitem{vandalen:2005b} E.N.E. van Dalen, C. Fuchs, and  A. Faessler, Phys. Rev. C \textbf{72}, 065803 (2005).
%
\bibitem{ramos:1990} A. Ramos, A. Polls, and W. H. Dickhoff, Nucl. Phys. \textbf{A503}, 1 (1990).
%
\bibitem{jaminon:1989} M. Jaminon and C. Mahaux, Phys. Rev. C \textbf{40}, 354 (1989).
%
\bibitem{gogelein:2009} P. G\"{o}gelein, E.N.E. van Dalen, Kh. Gad, Kh. S. A. Hassaneen, and H. M\"{u}ther, Phys. Rev. C \textbf{79}, 024308 (2009).
%
\bibitem{mahaux:1985} C. Mahaux, P.F. Bortignon, R.A. Broglia, and C.H. Dasso, Phys. Rep. \textbf{120},  1 (1985).
%
\end{thebibliography}
\end{document}